\documentclass[final,1p,times]{elsarticle} 
\usepackage{graphicx} 
\usepackage{amssymb} 
\usepackage{amsthm} 
\usepackage{lineno} 

\journal{Nuclear Physics A} 
\begin{document} 

\begin{frontmatter} 


\title{Describing transverse dynamics and space-time evolution at RHIC in
a hydrodynamic model with statistical hadronization}

\author{W. Florkowski$^{a,b}$, W. Broniowski$^{a,b}$, M. Chojnacki$^b$, A.~Kisiel$^{c,d}$}

\address[a]{Institute of Physics, Jan~Kochanowski University, 25-406 Kielce, Poland}
\address[b]{The H.~Niewodnicza\'nski Institute of Nuclear Physics, 
Polish Academy of Sciences, 31-342 Krak\'ow, Poland}
\address[c]{Faculty of Physics, Warsaw University of Technology, 00-661 Warsaw, Poland}
\address[d]{Physics Department, CERN, CH-1211 Geneve 23, Switzerland}

\begin{abstract} 
A hydrodynamic model coupled to the statistical hadronization code Therminator is used to study a set of observables in the soft sector at RHIC. A satisfactory description of the \mbox{$p_\perp$}-spectra and elliptic flow is obtained, similarly to other hydrodynamic models. With the Gaussian initial conditions the transverse femtoscopic radii are also reproduced, providing a possible solution of the RHIC HBT puzzle.
\end{abstract} 

\end{frontmatter} 



\section{Intoduction}

The consistent description of various features of the soft hadron production in the nucleus-nucleus collisions at the Relativistic Heavy-Ion Collider (RHIC) is a well known problem \cite{Heinz:2002un}. The so called RHIC HBT puzzle \cite{Heinz:2002un,Hirano:2004ta,Lisa:2005dd,Huovinen:2006jp} refers to the difficulty of simultaneous description of the hadronic transverse-momentum spectra, the elliptic flow coefficient $v_2$, and the Hanbury-Brown--Twiss (HBT) interferometry data in various approaches including hydrodynamics \cite{Heinz:2001xi,Hirano:2001yi,Hirano:2002hv,Zschiesche:2001dx,Socolowski:2004hw}. 

Recently, we have constructed a hydrodynamic code \cite{Chojnacki:2006tv,Chojnacki:2007rq} that is coupled to the statistical hadronization model Therminator \cite{Kisiel:2005hn}. Within this approach a successful uniform description of the soft hadronic RHIC data has been accomplished \cite{Broniowski:2008vp}. The main ingredients of our approach are the following: {\bf i)} We use a realistic equation of state that interpolates between the lattice QCD results at high temperatures and the hadron-gas results at lower temperatures \cite{Chojnacki:2007jc}. {\bf ii)} The single freeze-out scenario including all well established resonance states is assumed \cite{Broniowski:2001we}. This assumption leads to shorter emission times and helps to reproduce the ratio $R_{\rm out}/R_{\rm side}$. {\bf iii)} The use of Therminator  allows for the use of two-particle methods in the evaluation of the correlation functions (with and without the Coulomb corrections). {\bf iv)} The early starting/thermalization time \mbox{$\tau_{\rm i} = 0.25$ fm} is assumed. This helps to develop fast the transverse flow and shortens the evolution time. {\bf v)} The initial conditions for the energy density in the transverse plane are taken in the Gaussian form \cite{Broniowski:2008vp}
\begin{eqnarray}
  \varepsilon({\bf x}_\bot) = \varepsilon_{\rm i} \exp \left ( -\frac{x^2}{2a^2} -\frac{y^2}{2 b^2} \right ).
  \label{eqn:resrhic2_e0}
\end{eqnarray} 
The values of the width parameters $a$ and $b$ depend on the centrality class The Gaussian profiles are obtained with the Monte-Carlo Glauber simulations done with Glissando \cite{Broniowski:2007nz}. {\bf vi)} The initial conditions include fluctuations of the initial eccentricity \cite{Socolowski:2004hw,Voloshin:2008dg}.

\section{Results}

\begin{figure}[ht]
\centering
\includegraphics[width=0.5\textwidth]{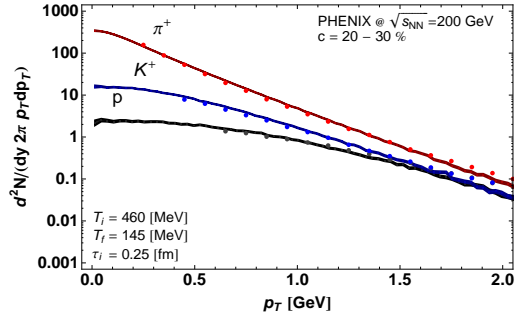}
\caption[]{The transverse-momentum spectra of pions, kaons and protons for the centrality class 20-30\%. The model results with the Gaussian initial conditions are plotted as functions of the transverse momentum and compared to the data from \cite{Adler:2003cb}.}
\label{fig:rhicTi460c2030-sppt}
\end{figure}

In Fig.~\ref{fig:rhicTi460c2030-sppt} we show our model results for the transverse-momentum spectra and compare them with the PHENIX data \cite{Adler:2003cb} for Au+Au collisions at the top RHIC energy of $\sqrt{s_{NN}}=200$~GeV. One observes a very good agreement between the model predictions and the data. The model results have been obtained with the initial central temperature \mbox{$T_{\rm i} =$ 460 MeV} and the final/freeze-out temperature \mbox{$T_{\rm f} =$ 145 MeV}. These are the two main parameters of our approach. The values of the width parameters follow from the Monte-Carlo Glauber simulations and for the centrality class 20-30\% we have found $a=2.00$~fm and $b=2.59$~fm. 

\begin{figure}[ht]
\centering
\includegraphics[width=0.5\textwidth]{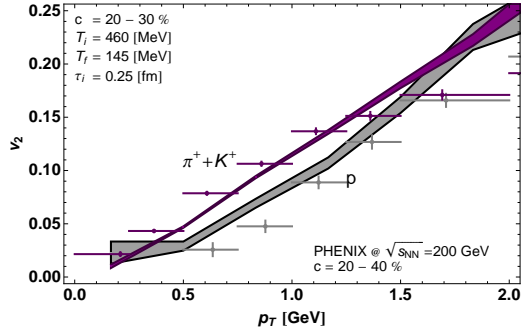}
\caption[]{The elliptic flow coefficient $v_2$ for the centrality class 20-40\%. The model results with the Gaussian initial conditions are plotted as functions of the transverse momentum and compared to the data from \cite{Adler:2003kt}. Model parameters are the same as in Fig.~\ref{fig:rhicTi460c2030-sppt}.}
\label{fig:rhicTi460c2030-v2}
\end{figure}

In Fig.~\ref{fig:rhicTi460c2030-v2} we show our model results for the elliptic flow coefficient $v_2$. The same values of the parameters have been used as in the calculation of the spectra. The pion+kaon data are very well reproduced, while the model results for the protons slightly overshoot the data. This behavior may be attributed to the lack of hadron rescattering in the final state in our approach.

\begin{figure}[h]
  \begin{center}
  \includegraphics[width=0.5\textwidth]{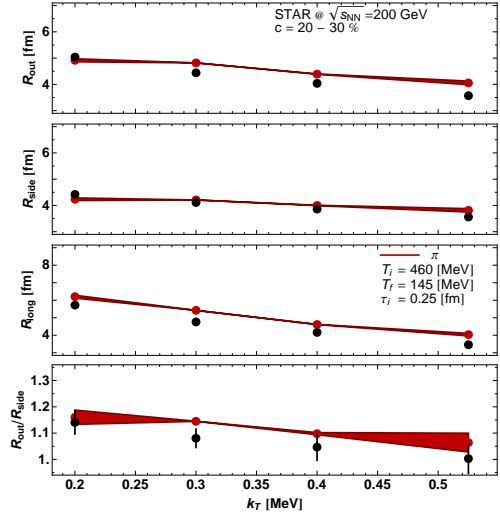}
  \end{center}
  \caption{\small The pion HBT radii $R_{\rm side}$ , $R_{\rm out}$ , $R_{\rm long}$, and the ratio $R_{\rm out}/R_{\rm side}$ for the centrality class 20-30\%. The results of the model calculation with the Gaussian initial conditions (lines) are compared to the data from \protect\cite{Adams:2004yc} (dots). Model parameters are the same as in Fig.~\ref{fig:rhicTi460c2030-sppt}.}
  \label{fig:resrhic2nofs-hbt}
\end{figure}

Our HBT results are shown in Fig.~\ref{fig:resrhic2nofs-hbt}. Again, the same values of the input parameters have been used. We observe good agreement between the data \cite{Adams:2004yc} and the model calculations. In particular, the ratio $R_{\rm out}/R_{\rm side}$ is very well reproduced. 

\begin{figure}[h]
\begin{center}
\includegraphics[angle=0,width=0.95 \textwidth]{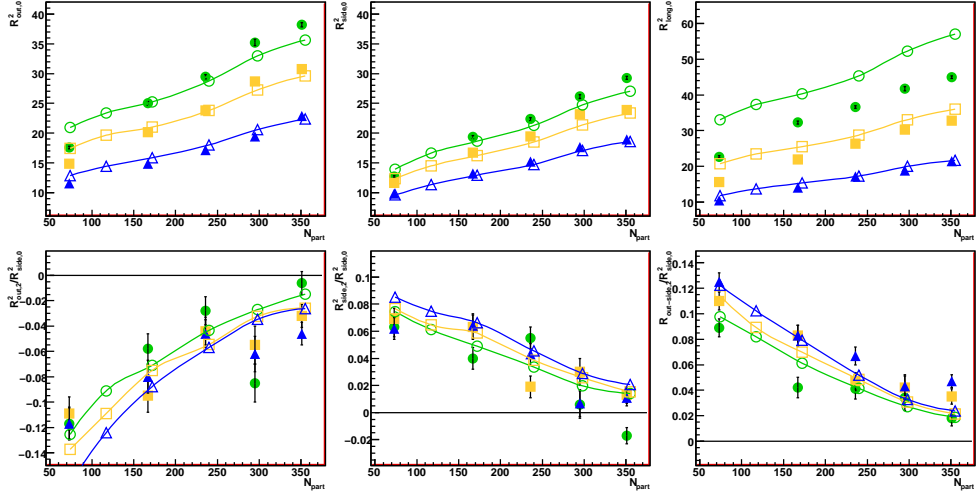}
\end{center}
\vspace{-6.5mm}
\caption{\small Results for the RHIC HBT radii and their azimuthal oscillations. For each value of $N_{\rm part}$ on the horizontal axis we plot the experimental points (filled symbols) and the model results (empty symbols). The points from top to bottom at each plot correspond to $k_T$ contained in the bins 0.15-0.25~GeV (circles), 0.25-0.35~GeV  (squares), and 0.35-0.6~GeV(triangles). The top panels show $R^{2}_{\rm out,0}$, $R^{2}_{\rm side,0}$, and $R^{2}_{\rm long,0}$, the bottom panels the magnitude of the allowed oscillations divided conventionally by $R_{\rm side,0}^2$. Data from Ref.~\cite{Adams:2003ra}. Model parameters for different centralities are given in \cite{Kisiel:2008ws}.}
\label{fig:resultsrhic}
\end{figure}

Encouraged by the success of reproducing the azimuthally averaged HBT radii, we have also calculated the azimuthally sensitive femtoscopic observables for different centralities and average transverse momenta \cite{Kisiel:2008ws}. The summary of our results is shown in Fig.~\ref{fig:resultsrhic}. The model results are compared to the experimental STAR data \cite{Adams:2003ra}. For each centrality, associated here with the number of participants $N_{\rm part}$ on the horizontal axis, we plot the experimental points (filled dots) and the model results (empty symbols). The points from top to bottom correspond to $k_T$ contained in the bins of 0.15-0.25~GeV, 0.25-0.35~GeV, and 0.35-0.6~GeV. The top panels show the radii squared averaged over the $\phi$ angle, from left to right, $R^{2}_{\rm out,0}$, $R^{2}_{\rm side,0}$, and $R^{2}_{\rm long,0}$. The bottom panels show the magnitude of the allowed oscillations divided by $R_{\rm side,0}^2$, which is the adopted convention used in presenting the experimental data. 

We conclude with the statement that the consistent and uniform description of the soft hadronic data at RHIC may be achieved within the hydrodynamic approach if a proper choice of the initial profile and a realistic equation of state are used. We note that a similar conclusion has been also reached recently in Ref. \cite{Pratt:2008qv} (see also \cite{Lisa:2008gf}) where the fast building of the transverse flow that is required for the correct description of the HBT radii is achieved with the very early start of the hydrodynamics ($\tau_{\rm i}$ = 0.1 fm) and the inclusion of the viscous effects. However, the $v_2$ coefficient is not evaluated in \cite{Pratt:2008qv}.


\section*{Acknowledgments} 
This work was supported in part by the Polish Ministry of Science and Higher Education, 
grant N202 034 32/0918.

\end{document}